%% file: paper.tex
\documentclass[usenatbib,usegraphicx,epsfig]{aa}
\usepackage{graphicx}
\usepackage{txfonts}

\usepackage{supertabular,multicol}
\newcommand{\mgii}{\ion{Mg}{ii}~}
\newcommand{\civ}{\ion{C}{iv}~}

\begin{document}

   \title{Solving the conundrum of intervening strong \mgii absorbers
     towards GRBs and quasars}

   \author{L. Christensen\inst{1}, S.~D.~Vergani\inst{2,3},
     S.~Schulze\inst{4}, N.~Annau\inst{5,1}, J.~Selsing\inst{1},
     J.~P.~U.~Fynbo\inst{1},
     A.~de~Ugarte~Postigo\inst{6,1},
     R.~Ca\~{n}ameras\inst{1}, 
     S.~Lopez\inst{7}, D.~Passi\inst{7},
     P.~Cort\'es-Zuleta\inst{7},
     S.~L.~Ellison\inst{5},
     V.~D'Odorico\inst{8},  
     G.~Becker\inst{9},
     T.~A.~M.~Berg\inst{5},
     Z.~Cano\inst{6},
     S.~Covino\inst{10},
     G.~Cupani\inst{8},
     V.~D'Elia\inst{11,12},
     P.~Goldoni\inst{13},
     A.~Gomboc\inst{14},
     F.~Hammer\inst{2},
     K.~E.~Heintz\inst{15,1},
     P.~Jakobsson\inst{15},
     J.~Japelj\inst{16},
     L.~Kaper\inst{16},
     D.~Malesani\inst{1},
     P.~M{\o}ller\inst{17},
     P.~Petitjean\inst{3},
     V.~Pugliese\inst{16},
     R.~S\'anchez-Ram\'irez\inst{6,18},
     N.~R.~Tanvir\inst{19},
     C.~C.~Th{\"o}ne\inst{6},
     M.~Vestergaard\inst{1,21},
     K.~Wiersema\inst{19},
     G.~Worseck\inst{20}
     \thanks{Based on observations
       collected at the European Southern Observatory, Paranal, Chile,
       Program ID: 098.A-0055, 097.A-0036, 096.A-0079, 095.B-0811(B),
       095.A-0045, 094.A-0134, 093.A-0069, 092.A-0124, 0091.C-0934,
       090.A-0088, 089.A-0067, 088.A-0051, 087.A-0055, 086.A-0073,
       085.A-0009 and 084.A-0260. XQ-100: 189.A-0424.}
   }
   \institute{Dark Cosmology Centre, Niels Bohr Institute,
  University of Copenhagen, Juliane Maries Vej 30, DK-2100 Copenhagen,
  Denmark  \email{\tt lise@dark-cosmology.dk}
  \and GEPI, Observatoire de Paris, PSL Research University, CNRS,
Place Jules Janssen, 92190 Meudon, France
   \and Institut d'Astrophysique de Paris, Universit{\'e} Paris
   6-CNRS, UMR7095, 98bis Boulevard Arago, F-75014 Paris, France
   \and Department of Particle Physics and Astrophysics, Weizmann Institute of Science, Rehovot 7610001, Israel
  \and Department of Physics and Astronomy, University of Victoria,
  Victoria, BC V8P 1A1, Canada 
  \and Instituto de Astrof\'{\i}sica de Andaluc\'{\i}a (IAA-CSIC),
  Glorieta de la Astronom\'{\i}a s/n, E-18008, Granada, Spain
  \and Departamento de Astronom\'ia, Universidad de Chile, Casilla 36-D, Santiago
  \and INAF- Osservatorio Astronomico di Trieste, Via Tiepolo 11, I-34143 Trieste, Italy
  \and Department of Physics and Astronomy, University of California, Riverside, CA, 92521, USA
   \and INAF / Osservatorio Astronomico di Brera, via Bianchi 46,
   23807 Merate (LC), Italy
   \and INAF-Osservatorio Astronomico di Roma, Via Frascati 33, I-00040 Monteporzio Catone, Italy
   \and ASI-Science Data Centre, Via del Politecnico snc, I-00133 Rome, Italy
   \and APC, Univ. Paris Diderot, CNRS/IN2P3, CEA/Irfu, Obs. de Paris, Sorbonne Paris Cit\'e, 75013 P
   \and Centre for Astrophysics and Cosmology, University of Nova
   Gorica, Vipavska 11c, 5270 Ajdov{\v s}{\v c}ina, Slovenia
   \and Centre for Astrophysics and Cosmology, Science Institute, University of Iceland, Dunhagi 5, 107 Reykjav\'ik, Iceland
   \and Anton Pannekoek Institute for Astronomy, University of
   Amsterdam, Science Park 904, 1098 XH Amsterdam, The Netherlands
   \and European Southern Observatory, Karl-Schwarzschild-Strasse 2,
   D-85748 Garching bei M{\"u}nchen, Germany
   \and INAF, Istituto di Astrofisica e Planetologia Spaziali, Via
   Fosso del Cavaliere 100, I-00133 Roma, Italy
   \and Department of Physics and Astronomy, University of Leicester, Leicester LE1 7RH, UK
   \and Max-Planck-Institut f\"ur Astronomie, K\"onigstuhl 17, D-69117
   Heidelberg, Germany
   \and Dept. of Astronomy, Steward Observatory, University of Arizona, 933 North Cherry Avenue, Tucson, AZ 85721
   }
   \date{Received 15 June 2017; Accepted 1 September 2017 }  


\authorrunning{L. Christensen et al.}
\titlerunning{Intervening strong \mgii absorbers}

 \abstract{Previous studies have shown that the incidence rate of
   intervening strong \mgii absorbers towards GRBs were a factor of
   $2-4$ higher than towards quasars. Exploring the similar sized and
   uniformly selected legacy data sets XQ-100 and XSGRB, each
   consisting of 100 quasar and 81 GRB afterglow spectra obtained with a
   single instrument (VLT/X-shooter), we demonstrate that there is no
   disagreement in the number density of strong \mgii absorbers with
   rest-frame equivalent widths $W_r^{\lambda2796}>1$ {\AA} towards
   GRBs and quasars in the redshift range $0.1\lesssim z\lesssim 5$.
   With large and similar sample sizes, and path length coverages of
   $\Delta z=57.8$ and $254.4$ for GRBs and quasars, respectively, the
   incidences of intervening absorbers are consistent within 1$\sigma$
   uncertainty levels at all redshifts. For absorbers at $z<2.3$ the
   incidence towards GRBs is a factor of $1.5\pm0.4$ higher than the
   expected number of strong \mgii absorbers in SDSS quasar spectra,
   while for quasar absorbers observed with X-shooter we find an excess
   factor of $1.4\pm0.2$ relative to SDSS quasars. Conversely, the
   incidence rates agree at all redshifts with reported high spectral
   resolution quasar data, and no excess is found.  The only remaining
   discrepancy in incidences is between SDSS \mgii catalogues and high
   spectral resolution studies.  The rest-frame equivalent width
   distribution also agrees to within 1$\sigma$ uncertainty levels
   between the GRB and quasar samples. Intervening strong \mgii absorbers
   towards GRBs are therefore neither unusually frequent, nor
   unusually strong. }

    \maketitle

   %

\section{Introduction}
Luminous point sources like gamma-ray bursts (GRBs) and quasars are
efficient probes of the gaseous material along their lines of
sight. Both classes of objects are currently detected out to $z>7$
\citep{tanvir09,salvaterra09,Mortlock11} and thus probe a very long
path length through the universe.  Absorption lines at various
redshifts in the spectra of the background sources provide us with
methods to explore the high-redshift universe in absorption even
though the galaxies that cause the absorption lines are not detected
in emission. Since both GRBs and quasars probe intervening material
randomly, it was a puzzling discovery a decade ago that GRBs
apparently had four times as many strong intervening \mgii absorbers
with rest-frame equivalent widths $W_r^{\lambda2796} > 1$~{\AA} as did
quasars \citep{prochter06g}. The same conclusion about \mgii absorber
statistics was reached by \citet{sudilovsky07}, with the addition that
the incidence rate of \civ absorbers and weak \mgii absorbers with
$W_r < 1$ {\AA} did agree between the two background source types
\citep{tejos07,tejos09,vergani09}.  Also blazars were found to have
twice as many strong \mgii absorbers as did quasars
\citep{bergeron11}. Exploring high spectral resolution VLT/UVES data
of GRB afterglows with high signal-to-noise (S/N) levels,
\citet{vergani09} and \citet{tejos09} found an excess, albeit only of
a factor of 2--3 higher towards GRBs than Sloan Digital Sky Survey
(SDSS) quasars, and with only a 2-$\sigma$ confidence level of the
over-abundance. Table~\ref{tab:literature} presents a summary of
previous searches and reported excesses of strong \mgii systems.  As
the discrepancy between the absorber statistics has only been reported
for strong \mgii absorbers with $W_r^{\lambda2796} > 1$~{\AA}, in this
work we will focus exclusively on these strong systems, unless
specifically stated otherwise.

The excess number of absorbers has been suggested to be caused by
either dust biases, partial covering and differences in source sizes,
gravitational magnification, or the influences from the immediate
environments of the GRBs, or intrinsic to the sources
\citep{prochter06g,porciani07,frank07,menard08,cucchiara09,vergani09,budzynski11,rapoport13}. Any
effect of a partial coverage and source sizes was excluded based on
statistics of absorbers towards the quasar broad line regions
\citep{pontzen07,lawther12}. Some studies were hampered by relatively
small GRB afterglow sample sizes and non-uniform data
sets. \citet{cucchiara13} compiled a large data-set collected from
various telescopes and with a range of spectral resolutions, and
demonstrated that the low- to intermediate-resolution data showed
consistent values of strong \mgii absorbers between GRBs and
quasars. However, when including the original high-resolution data
from \citet{prochter06g} the discrepancy remained.

\begin{table*}
  \centering
  \begin{tabular}{lcccccc}
\hline
\hline
Reference &  $N_{\mathrm{obj}}$  &  $N_{\mgii}$ & $\Delta z$ & $\langle z
\rangle$ & excess/SDSS & excess/High-res  \\
\hline
\multicolumn{7}{c}{Low redshift absorbers  ($z_{\mgii}<2.3$)}\\[1ex] 
\cite{prochter06g}          & 14 & 14 &   ~15.5 & 1.1  & $\approx3.8$ \\
\cite{sudilovsky07}         &  5 &  6 & ~~~6.75 & 1.3  & $\approx4$ \\
\cite{tejos09}              &  8 &  9 &   10.86 & 1.34 & $3.0^{+1.5}_{-1.1}$\\
\cite{bergeron11} (blazars) & 45 & 18 &   25.11 & 0.82 & $2.2^{+0.8}_{-0.6}$\\
\cite{vergani09} (high res.) & 10 &  9 &   13.94 & 1.11 & $\approx2$ \\
\cite{vergani09} (high+low res.) & 26 & 22 & 31.55 & 1.3 & $2.1\pm0.6$ \\
\cite{cucchiara13} (high res.) & 18 & 13 & 20.3 & 1.1    & $2.6\pm0.8$\\
\cite{cucchiara13} (high+low res.) & 95 & 20 & 55.5 & 1.15 & $1.5\pm0.4$\\
This work (GRBs from XSGRB)  & 81 & 18 & 44.71 & 1.22 & $1.48\pm0.35$  & $1.11\pm0.28$\\
This work (quasars)  & 100 & 52 & 110.5 & 1.74 & $1.37\pm0.19$  & $1.02\pm0.18$\\[1ex] 
\hline
\hline
\multicolumn{7}{c}{High redshift absorbers ($z_{\mgii}>2.3$)}\\[1ex] 
This work (GRBs from XSGRB)  & 81 & 5 & 13.14 & 3.28 &
  & $0.71\pm0.34$\\
This work (quasars)  & 100 & 45 & 143.9 & 3.04 &
  & $0.56\pm0.93$\\
\hline
\end{tabular}
    \caption{Strong \mgii absorber searches in the
      literature. $N_{\mathrm{obj}}$ lists the GRB or quasar sample
      sizes, and $N_{\mgii}$ is the number of detected absorbers
      within the total redshift path length, $\Delta z$, with an
      average redshift, $\langle z \rangle$. The penultimate column
      lists the reported excess of absorbers relative to the
      expectation from SDSS quasar absorber statistics. One exception
      is that \citet{bergeron11} analysed blazars rather than
      GRBs. \citet{vergani09} and \citet{cucchiara13} reported
      statistics from high-spectral resolution data, as well as
      combined high- and low spectral resolution data. It should be
      noted that there are substantial overlaps of the target
      selections amongst the references, and the reported
      overdensities are therefore correlated. To compare our searches
      with the literature, we report statistics from absorbers at
      different redshift ranges: $z<2.3$ and $z>2.3$
      (Sect.~\ref{sect:comp}). The last column reports overdensities
      relative to high spectral resolution quasar studies by
      \citet{mathes17} for the low-redshift sample, and \citet{chen16}
      for the high-redshift sample.}
  \label{tab:literature}
\end{table*}

Incidences ($dN/dz$, or sometimes referred to as $l(z)$) of \mgii
systems in GRB spectra are commonly compared to strong \mgii absorbers
in SDSS quasar spectra, where absorber identifications are performed by
automated routines \citep{nestor05,prochter06q,zhu13}. In SDSS data
release DR12, the number of SDSS quasar spectra \citep[$\sim$300,000
  in][]{paris17} vastly outnumber GRB afterglow spectra
($\sim$300\footnote{{\tt grbspec.iaa.es} presents afterglow spectra
  for 225 GRBs compiled until June, 2017 \citep{deugarte14}.})
\citep{fynbo09,deugarte12}, and so do the number of intervening \mgii
absorbers \citep[$\sim$37,000 in][]{raghunathan16}.

In this paper we revisit the \mgii puzzle by exploring newly collected
homogeneous data sets obtained with a single instrument, X-shooter/VLT
\citep{vernet11}. Data of 100 quasars were taken from the Legacy Large
Programme, XQ-100 \citep{lopez16}, and 116 GRB spectra from the
X-Shooter GRB legacy sample (XSGRB; Selsing et al. 2017 in prep, PI:
Fynbo). These legacy data sets allow us to explore two unbiased and
equally large sample sizes all obtained with a uniform spectral
resolution and instrument setup. With these data, we expand the
comparison of the \mgii absorber incidence to a larger redshift
interval $0.14<z<5$. Section \ref{sect:data} describes the two data
sets (GRB afterglows and quasars) and the search for strong absorbers. In
Sect.~\ref{sect:ldens} we derive incidence rates, equivalent width
distributions and compare with other studies in the literature, and
then summarise in Sect.~\ref{sect:conc}.


\section{Data sets and \mgii absorber detection}
\label{sect:data}

\subsection{X-shooter spectral samples}
The GRB afterglow spectra were collected in multiple semesters between
2009 and 2016 taking advantage of the X-shooter guaranteed program and
subsequent open time proposals. With a uniform set of selection
criteria for target-of-opportunity follow-up adopting the criteria for
the low-resolution study of \citet{fynbo09}, i.e. mainly avoiding
regions with high Galactic extinction and bursts near to the Sun, the
total number of obtained afterglow spectra is 116 until October,
2016. All spectra were obtained with VLT/X-shooter, which provides
intermediate spectral resolutions of $R=FWHM/\lambda
\approx6,500-12,000$ measured from UV to near-IR wavelengths.  The S/N
ratio of the spectra vary as some afterglow spectra were obtained
several hours to a few days after the GRB trigger, when only host
spectral signatures and no characteristic power-law afterglow
signature was visible.  Removing these from the sample leaves 81
afterglow spectra, again with varying S/N between 1--50 per pixel for
different bursts and wavelength coverage (see Selsing et al. 2017 in
prep for details). The data were reduced with the ESO VLT/X-shooter
pipeline \citep{modigliani10} managed by Reflex \citep{freudling13},
and with our own post-processing steps to improve the flux calibration
and rejection of bad pixels (Selsing et al. 2017 in prep) with
reduction scripts made available
online\footnote{\url{https://github.com/jselsing/XSGRB_reduction_scripts}}.
The redshifts of the GRBs lie in the range $0.256<z<6.32$.

Turning to the quasar sample, the XQ-100 survey is a VLT/X-shooter ESO
Large Programme that targeted 100 quasars at $3.5< z< 4.9$ with a S/N
ratio higher than 20 per pixel over the entire wavelength range
\citep[see][for details]{lopez16}. The quasars were selected without
prior knowledge of the presence of strong absorbers along our line of
sight. The quasars and a large majority of GRB afterglows were
observed with X-shooter slit widths of 1.0, 0.9, and 0.9 arcsec for
the UVB, VIS, and NIR-arms, respectively, resulting in similar
spectral resolutions between the two samples.

\begin{figure}[t!]
\centering
\includegraphics[bb=95 373 534 695, clip, width=8.6cm]{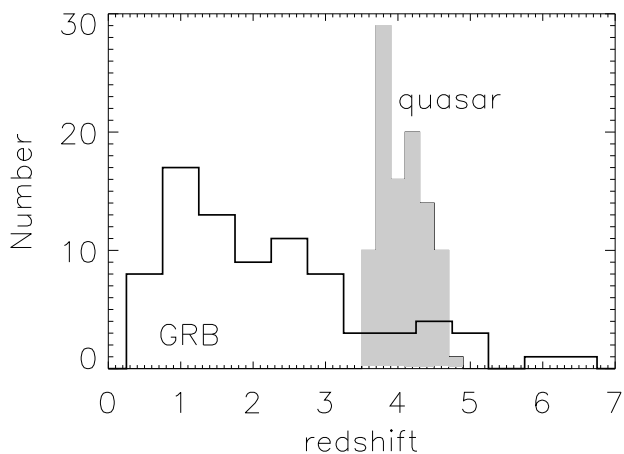}
 \caption{Redshift distributions for the two background source samples
   (XQ-100 and XSGRB).  The 81 GRB afterglows and 100 quasars have median
   redshifts of $z=1.69$ and $z=3.97$, respectively.}
\label{fig:z_dist}
\end{figure}

The redshift distributions of the GRB and quasar samples are shown in
Fig.~\ref{fig:z_dist}. By construction the quasars have a quite
distinct distribution as high redshift objects were selected, whereas
there are no intrinsic constraints for the GRB redshift selection.  To
match the redshift distribution of GRBs we could include other quasars
observed with X-shooter, using archive data. However, these quasars
might have been selected  on the basis of the presence or lack of
  any kind of intervening and intrinsic absorption line systems, so
we do not expand the quasar sample beyond the XQ-100 data.

\subsection{Additional data sets}
Since the earlier discrepancy on the incidence of \mgii absorbers
relied on the statistics of GRB absorbers obtained to that date, we
also examined another large spectroscopic sample consisting of 60
low-resolution GRB afterglow spectra with known redshifts observed
from 2005 to 2008. Intervening absorbers and their observed equivalent
widths were compiled by \citet{fynbo09}. These low-resolution spectra
are also included in the \mgii statistics analysis by
\citet{cucchiara13}. In this work, the low-resolution data-set is used
as a comparison sample to the XSGRB data.

\subsection{Redshift path length}
The starting point of computing the strong \mgii absorber incidences
is to determine the redshift path length
\citep[e.g.][]{lanzetta87}. Rather than relying on the sample sizes
alone, the most relevant parameter for a survey is the total redshift
path length where an absorber can be found.

X-shooter observes wavelengths down to 3000~{\AA}, but because of a
much reduced transmission below 3200~{\AA}, we chose this latter
wavelength to represent the lowest redshift for the detection of \mgii
absorbers ($z=0.14$).

For each object we created a top-hat function which was set to one
between the redshifted Lyman-$\alpha$ line of the GRB/quasar and the
wavelength of \mgii at the GRB/quasar redshift.  The lower redshift limit
excluded absorbers found in the Ly$\alpha$ forest because they are
likely contaminated by intervening Ly$\alpha$ forest lines. We also
excluded the region around 3000 km~s$^{-1}$ from the GRB/quasar
redshifts, because absorbers detected in this region may be associated
with the luminous background sources. These proximate absorbers have
higher metallicities \citep{ellison11} and different incidences
\citep{ellison02} showing that they do not probe a random intervening
population. Rejecting the nearby regions decreased the redshift path
lengths by $\Delta z =0.02-0.07$ for each individual line of
sight. Furthermore, we considered it impossible to find strong
intervening absorbers in regions heavily affected by strong telluric
lines when the transmission was less than $\sim$30\%, such as present
between the $J$- and $H$, and $H$- and $K$ bands. In these
inaccessible regions the function was set to zero. Other wavelength
regions are also affected by telluric absorption lines, but these
absorption lines have been corrected for \citep[][and Selsing et
  al. in prep. 2017]{lopez16}, and do not pose a problem for detecting
strong \mgii absorbers because their lines are much broader than the
widths of telluric molecular absorption lines.

We also imposed a criterion for the spectral S/N ratio. For the GRB
spectra, we set the function to zero in any region where the S/N ratio
was less than three per spectral pixel \citep[see][for a similar
  approach]{cucchiara13}.  A threshold of S/N=3 corresponds to a
3$\sigma$ detection limit \(W_{r,lim}^{\lambda2796}\approx
\sqrt{n_{pix}}\times2796{\AA}/R \approx 1~{\AA}\), where $R$ is the
spectral resolution and $n_{pix}$ is the number of pixels spanned by
the absorption line. Resolutions were determined to be $R\sim12,000$
and $R\sim6500$ from measuring the FWHM of unresolved lines in the VIS
and NIR arms, respectively (Selsing et al. in prep).

\begin{figure}[t!]
\centering
\includegraphics[bb=85 370 534 695, clip,width=8.5cm]{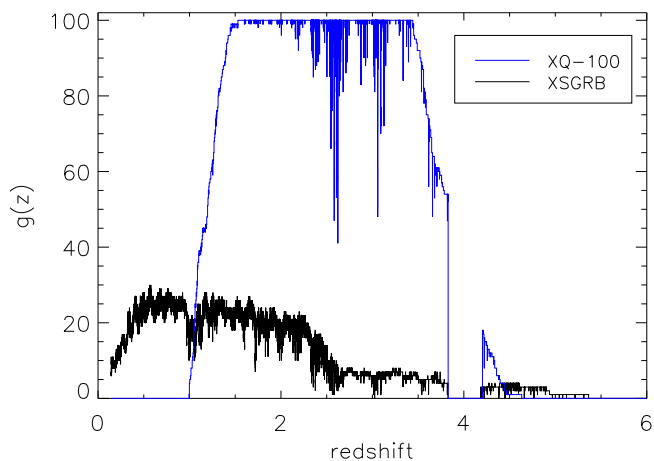}
 \caption{Redshift path density for the two samples, with quasars shown
   in blue and GRB afterglows in black. The gap at $z\sim4$ is caused
   by the atmospheric absorption bands between the $J$- and $H$ bands,
   where no \mgii absorbers can be recognised. XQ-100 includes 100
   sightlines, hence $g(z)$ saturates at this level. }
\label{fig:obj_dens}
\end{figure}

We then added up these functions for all the GRB and quasar spectra,
converted the observed wavelengths to the \mgii redshift values, and
summed over all objects as a function of the absorber redshift. The
resulting redshift paths densities, $g(z)$, are presented in
Fig.~\ref{fig:obj_dens}, and the integrated redshift path lengths \(
\Delta z=\int g(z)dz \) are listed in Table~\ref{tab:literature}.

\subsection{Finding \mgii absorbers}

As the discrepancy between \mgii absorber incidences refers to strong
systems, we searched for absorbers with $W_r^{\lambda2796}>1~{\AA}$.
Other species of atomic absorption lines, including lower equivalent
width \mgii systems, will be presented elsewhere (For the XQ-100
sample in S. Lopez et al. in prep. and the GRB sample in de Ugarte
Postigo et al. in prep.).

Strong \mgii lines with rest frame equivalent widths
$W_r^{\lambda2796} > 1~{\AA}$ were independently identified visually
by several of the authors. At the resolution of X-shooter, there is no
problem in identifying strong \mgii $\lambda\lambda$2796,2803 doublets
visually as the lines are well separated. Additional confirmation of
the doublet comes from identifying the transitions of \ion{Mg}{i}
$\lambda$2852 and \ion{Fe}{ii} $\lambda$2600 at the \mgii redshift,
but this criterion did not exclude any candidate \mgii systems.

With the chosen criterion that the S/N level is $>3$ per pixel, visual
identification of strong \mgii absorbers takes advantage of the
doublet nature with lines for the strong absorbers always being broad
and saturated.  The higher S/N ratio of the spectra in the XQ-100
sample meant that visual identification of strong \mgii absorbers was
unambiguous, and the choice of S/N cut did not affect the final
results. We verified that none of the identified strong absorbers fall
in wavelength regions affected by any of the selection criteria.
While the S/N ratio per spectral pixel is modest for some spectra in
the GRB sample, the absorption signature from the \mgii doublet plus
additional absorption species covered many pixels (30–50), and
therefore the integrated S/N of the absorption system is much larger
than the simple pixel-by-pixel level of the signal.  In total we
identified 23 strong \mgii systems towards the 81 GRBs and 97 strong
\mgii systems towards the 100 quasars. They are listed in
Tables~\ref{tab:list} and \ref{tab:list_qso}, respectively. We
computed the rest frame equivalent widths of the strongest line in the
doublet, $W_r^{\lambda2796}$, by defining a continuum level region
around the identified line with $\sim$10~{\AA} wide wavelength ranges
blue- and redwards of the \mgii doublet. Errors for $W_r$ were
computed by propagating the uncertainties from the associated error
spectrum. In the case where two absorber components lie within 500
km~s$^{-1}$ from each other they were treated as a single system
\citep[similarly as in][]{chen16}, and $W_r$ represents the sum of the
components.

We excluded absorbers with $W_r^{\lambda2796}<1$ {\AA} even if their
equivalent width, including uncertainties, could place them among the
strong absorber sample. Since the quasar spectra have high S/N values,
the uncertainties lie in the range $\Delta W_r = 0.01-0.04$ {\AA}.
Although below the $W_r>1$ {\AA} limit, including 1$\sigma$
measurement uncertainties, five additional \mgii absorbers would pass
for strong absorbers in the quasar sample and a single one in the GRB
sample. These are also listed in Tables~\ref{tab:list} and
\ref{tab:list_qso} for completeness.

The transient nature of GRB afterglows prevents us from obtaining
follow-up spectra at different wavelengths once the afterglow fades
below detection limits. Therefore it is difficult to determine
additional absorption properties of the strong intervening \mgii
absorbers in the XSGRB sample, for example if the absorbers are
metal-rich or strong depending on the hydrogen column
densities. However, with alternative methods we can address this issue
for the \mgii absorbers. The damped Ly-$\alpha$ absorbers (DLAs)
identified in XQ-100 \citep{sanchez-ramirez16,berg16} all show \mgii
absorption lines (besides those that remain undetected in the gap
between the $J$- and $H$ bands), but only about half are classified
strong \mgii systems with $W_r^{\lambda2796}>1$ {\AA}
\citep{berg17}. The $D$-index of \citet{ellison06a}, defined as
\(D=(W_r/\Delta V) \times 1000\), with $\Delta V$ being the full
velocity width of the absorption line in km~s$^{-1}$ relative to the
continuum, can be used as a criterion to pre-select DLA systems in
cases where the hydrogen column density is unknown. The threshold
value of the $D$-index for an absorber to be a DLA system depends on
the observed spectral resolution. All DLAs found in the XQ-100 sample
have $D>4$ \citep{berg17}. In Table~\ref{tab:list}, we list the
$D$-index for \mgii absorbers in the XSGRB sample, suggesting that all
but one of the absorbers are DLAs.

\begin{table}
  \begin{tabular}{lllll}
\hline
\hline
GRB name &  $z_{\mathrm{GRB}}$ & $z_{\mathrm{abs}}$ & $W_r^{2796}$
[{\AA}]  & $D$\\
\hline
GRB~090313A &     3.373  & 1.801  & 1.86$\pm$0.16           & 11.0$\pm$0.9\\
GRB~100219A &     4.667  & 1.856  & 1.02$\pm$0.08           & 5.9$\pm$0.5\\
            &            & 2.181  & 0.92$\pm$0.19           &10.5$\pm$2.2\\
GRB~100316B &     1.180  & 1.063  & 1.28$\pm$0.08           & 6.2$\pm$0.4\\
GRB~100901A &     1.408  & 1.315  & 1.35$\pm$0.28           & 2.1$\pm$0.4$^a$ \\
GRB~111008A &     4.990  & 4.610  & 4.25$\pm$0.09           & 6.9$\pm$0.1\\
GRB~111107A &     2.893  & 1.998  & 2.60$\pm$0.47           & 8.9$\pm$1.6\\
GRB~120119A &     1.728  & 1.214  & 1.65$\pm$0.08           & 4.2$\pm$0.2\\
GRB~120712A &     4.175  & 2.102  & 2.99$\pm$0.45           & 9.0$\pm$1.4\\
GRB~120815A &     2.359  & 1.539  & 6.26$\pm$0.07           & 9.3$\pm$0.1\\
GRB~121024A &     2.300  & 1.959  & 1.24$\pm$0.06           & 5.5$\pm$0.3\\
GRB~121027A &     1.773  & 1.459  & 1.91$\pm$0.21           & 9.5$\pm$1.0\\ 
GRB~121229A &     2.707  & 1.659  & 1.33$\pm$0.60           & 5.1$\pm$2.3\\
GRB~130408A &     3.758  & 3.016  & 2.96$\pm$0.83           & 8.0$\pm$2.3\\
GRB~130606A &     5.913  & 3.451  & 1.65$\pm$0.06           & 4.8$\pm$0.2\\
GRB~131030A &     1.294  & 1.164  & 1.95$\pm$0.03           & 4.9$\pm$0.1\\
GRB~140614A &     4.233  & 2.113  & 1.39$\pm$0.41           & 4.8$\pm$1.4\\
GRB~141028A &     2.332  & 1.820  & 3.23$\pm$0.73           & 9.9$\pm$2.2\\
GRB~141109A &     2.993  & 2.504  & 2.00$\pm$0.09           & 9.3$\pm$0.4\\
           &            & 2.874  & 4.21$\pm$0.25           & 10.1$\pm$0.6\\
GRB~150403A &     2.057  & 1.761  & 2.76$\pm$0.05           & 9.9$\pm$0.2\\
GRB~151021A &     2.330  & 1.491  & 1.35$\pm$0.33           & 7.1$\pm$1.7 \\
GRB~160203A &     3.519  & 1.267  & 1.40$\pm$0.19           & 4.0$\pm$0.5 \\
GRB~161023A &     2.708  & 1.243  & 1.83$\pm$0.02           & 7.1$\pm$0.1\\
\hline
\end{tabular}
    \caption{Intervening strong \mgii absorbers in the GRB sample,
      with redshifts listed for the GRBs ($z_{\mathrm{GRB}}$) and the
      absorbers ($z_{\mathrm{abs}}$). The list includes one absorber
      fulfilling the criterion that $W_r^{2796}~\gtrsim~1$~{\AA}
      within its 1$\sigma$ uncertainty.  The $D$ index refers to the
      definition by \citet{ellison06a}. $^a$This absorber has two
      distinct components separated by 390 km~s$^{-1}$ and the index
      refers to the sum of the components. }
  \label{tab:list}
\end{table}

In addition to the X-shooter samples, we examined the low resolution
GRB spectra by \citet{fynbo09} in the same manner as the X-shooter
data. This sample was kept separate from the X-shooter GRB sample, and
was used to look for any dependence on spectral resolution at lower
redshifts. In the low resolution sample we found 15 intervening strong
\mgii systems at $0.2<z<2.2$ within a total path length of $\Delta
z=50.3$.

\section{Statistics of strong \mgii systems}
\label{sect:ldens}

\subsection{Incidence rate}
Next we computed the incidence, or sometimes referred to as the line
density:
\begin{equation}
 \frac{dN}{dz} =\frac{\sum_{z_1}^{z_2}{N_{\mathrm{abs}}}}{\Delta z} =
  \frac{\sum_{z_1}^{z_2}{N_{\mathrm{abs}}}}{\int_{z_1}^{z_2} g(z)dz}  
\end{equation}
in redshift intervals from $z_1$ to $z_2$.  We experimented with
several binning approaches in the analysis, as described below.
Unless specified, in all experiments the first bin starts at the
minimum redshift in the path length, $z=0.14$. We first chose bin
sizes to include an equal number of absorbers in each bin, which gave
irregular sizes of the redshift bins. For the quasar sample, we chose
20 \mgii systems in each bin, six for the XSGRB sample, and eight in
the low-resolution GRB data. The numbers were chosen to have a similar
number of bins for each sample in order to analyse the redshift
evolution. With the small number of absorbers in the low-resolution
data, however, there were only two bins.

\begin{figure*}[htp!]
\centering
\begin{minipage}{0.5\textwidth}
\includegraphics[bb=75 85 495  678, clip,width=0.68\linewidth,angle=-90]{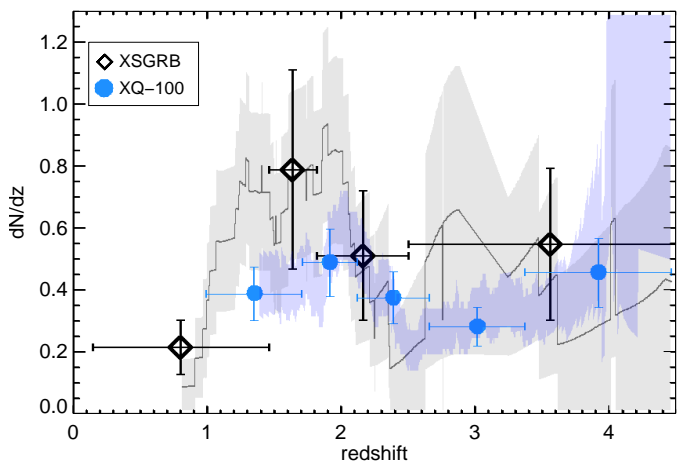}
\end{minipage}%
\begin{minipage}{0.5\textwidth}
\includegraphics[bb=75 373 534 700, clip,width=0.99\linewidth]{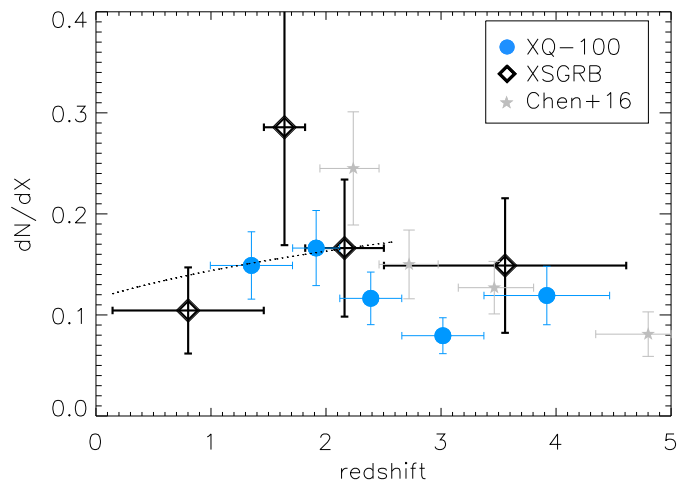}
\end{minipage}
 \caption{ {\it The left hand panel} shows the incidence rates from
   the two legacy samples of strong \mgii absorbers with points
   representing bin sizes with 6 and 20 absorbers in the XSGRB and
   XQ-100 samples, respectively.  The grey line illustrates a sliding
   redshift binning of XSGRB absorbers and the shaded region presents
   the 68\% confidence levels. The blue shaded region represents the
   68\% confidence intervals for a similar sliding redshift binning of
   the XQ-100 sample. {\it The right hand panel} shows the comoving
   line density. The dotted line represent strong \mgii absorbers in
   quasar spectra \citep{mathes17}, and the grey stars mark a larger
   sample of quasar absorbers at high redshifts \citep{chen16}.}
\label{fig:line_dens}
\end{figure*}

To compute uncertainties of the incidences we assumed that the error
is equal to $\sqrt{N_{\mathrm{abs}}}$ as given by Poisson
statistics. However, because the square root approximation
underestimates errors in the small number regime \citep{gehrels86}, we
also computed errors using a Monte Carlo bootstrapping with
replacement technique. We created a random sample with replacements of
100 quasars (81 GRBs) and compute the path lengths, number of absorbers
and incidences within the redshift intervals. The 1$\sigma$ standard
deviation of incidences from 1000 experiments, which represent the
uncertainty, gave the same error as computed from the square root
approximation for the Poisson statistics.

The results for the \mgii sample that passed the strict criterion that
$W_r > 1$~{\AA} are illustrated in the left panel in
Fig.~\ref{fig:line_dens} with computed incidences listed in
Table~\ref{tab:dndz}. We find that the incidence of strong \mgii
absorbers towards GRBs and quasars is consistent to within 1$\sigma$
uncertainty levels at all redshifts.

Traditional methods for computing the incidence also work with a set
of non-overlapping redshift bins. As a second method of binning, we
computed the incidence of \mgii systems in the GRB and quasar samples
using a sliding redshift bin technique \citep[see][for an application
  to compute the total neutral hydrogen density,
  $\Omega_{\mathrm{DLA}}$, with sliding bins]{sanchez-ramirez16}. We
chose redshift interval bin sizes of $\Delta z=0.7$ and increased the
steps above $z=2.5$ to $\Delta z=1.2$ for the GRB sample. The
resulting 68\% confidence intervals are shown in the grey and blue
shaded regions in the left panel in Fig.~\ref{fig:line_dens}. Choosing
a larger $\Delta z$ smooths out the curves, but the redshift evolution
of the incidence rate does not change the overall shape. We also
examined the result from using a sliding constant integrated path
length, $\Delta z$, \citep{sanchez-ramirez16}.  The two methods that
we used for computing the incidence rates gave consistent results.

The incidence rates were also computed with the expanded samples that
include absorbers with $W_r>1$ {\AA} within $\sim$1$\sigma$
measurement uncertainties. This addition does not change the results,
and the agreement between incidences from the GRB and quasar samples
remain consistent to within 1$\sigma$.

Another frequently used statistic related to absorbers is the comoving
line density $dN/dX$ defined as
\begin{equation}
  \frac{dN}{dX} = \frac{\sum_{z_1}^{z_2}{N_{\mathrm{abs}}}}{\Delta X},
  \end{equation}
where the absorption distance is
\begin{equation}
  \Delta X = \int_{z_1}^{z_2}
  g(z)\frac{(1+z)^2}{\sqrt{\Omega_m(1+z)^3+\Omega_{\Lambda}}} dz. 
\end{equation}
We used $\Omega_m=0.308$ and $\Omega_{\Lambda}=0.692$ from recent
Planck analyses \citep{planck16}. The results listed in
Table~\ref{tab:dndz} and shown in the right hand panel of
Fig.~\ref{fig:line_dens} again demonstrate that the quasar and GRB
incidence rates agree within 1$\sigma$ uncertainty levels.

The comoving line density of quasar \mgii absorbers has been argued to
evolve with redshift, with an increase from $z=0$ to $z\sim2$
\citep{nestor05,mathes17} followed by a decrease towards higher
redshifts \citep{matejek12,chen16}. This trend roughly follows the
redshift evolution of the comoving star-formation rate density,
suggesting that the strong absorbers somehow trace star-formation
activity \citep{menard11}. In the analyses of GRB and quasar absorbers
presented in this work, measurement uncertainties are large and do not
probe the highest redshifts ($z>5$). Nevertheless, number statistics
agree with previous high spectral resolution studies
\citep{matejek12,chen16}.

\begin{table}
  \begin{tabular}{cccc}
\hline
\hline
\multicolumn{4}{c}{GRB \mgii systems}\\[1ex]
$\langle z \rangle$ & $z_{\mathrm{min}} - z_{\mathrm{max}}$ &  $dN/dz$ & $dN/dX$ \\
\hline
0.802 & 0.144 -- 1.459  & 0.215$\pm$0.088 & 0.104$\pm$0.043 \\
1.639 & 1.459 -- 1.818  & 0.788$\pm$0.322 & 0.286$\pm$0.117 \\
2.161 & 1.818 -- 2.503  & 0.511$\pm$0.209 & 0.166$\pm$0.068 \\
3.557 & 2.503 -- 4.610  & 0.547$\pm$0.245 & 0.149$\pm$0.067 \\
\hline
\hline

\multicolumn{4}{c}{Quasar \mgii systems}\\[1ex]
$\langle z \rangle$ & $z_{\mathrm{min}} - z_{\mathrm{max}}$ &  $dN/dz$ & $dN/dX$ \\
\hline
1.352 & 0.963 -- 1.740 & 0.357$\pm$0.087 & 0.149$\pm$0.033 \\
1.914 & 1.740 -- 2.404 & 0.205$\pm$0.109 & 0.166$\pm$0.037 \\
2.389 & 2.404 -- 2.776 & 0.269$\pm$0.084 & 0.116$\pm$0.026 \\
3.016 & 2.776 -- 3.061 & 0.357$\pm$0.063 & 0.079$\pm$0.018 \\
3.920 & 3.061 -- 4.381 & 0.546$\pm$0.111 & 0.119$\pm$0.029 \\
\hline
\end{tabular}
    \caption{Incidences for strong \mgii absorbers towards GRB and
      quasars at the median redshift interval $\langle z \rangle$.
      $z_{\mathrm{min}} - z_{\mathrm{max}}$ gives the redshift
      interval.}
  \label{tab:dndz}
\end{table}

\subsection{Comparison between samples}
\label{sect:comp}
Earlier analyses suggested that GRBs and quasars trace \mgii systems
differently, and the excess incidence of strong \mgii absorbers are
frequently computed relative to SDSS quasar spectra. In order to
compare numbers with the overdensities listed in
Table~\ref{tab:literature}, we computed the expected number of strong
absorbers by integrating
\begin{equation}
  N_{\mathrm{exp}}= \int_{z_1}^{z_2} g(z) \frac{\partial N}{\partial z} dz
\label{eq:nabs}
\end{equation}
from $z_1=0.14$ to $z_2=2.3$ where the functional form for the SDSS
\mgii incidence $\frac{\partial N}{\partial z}$ was taken from
\citet{zhu13} and $g(z)$ from Fig.~\ref{fig:obj_dens}. The expected
  numbers of \mgii absorbers are 12.1 and 37.9 for the XSGRB and
  XQ-100 samples, respectively, while we found 18 and 52 absorbers.
  Both samples therefore suggest an excess factor of $1.48\pm0.35$ and
  $1.37\pm0.19$ relative to SDSS quasars, while the incidences in this
  redshift interval are consistent between our quasar and GRB samples
  within 0.3$\sigma$ uncertainties.

\begin{figure}
\includegraphics[bb=90 373 534 700, clip,width=0.99\linewidth]{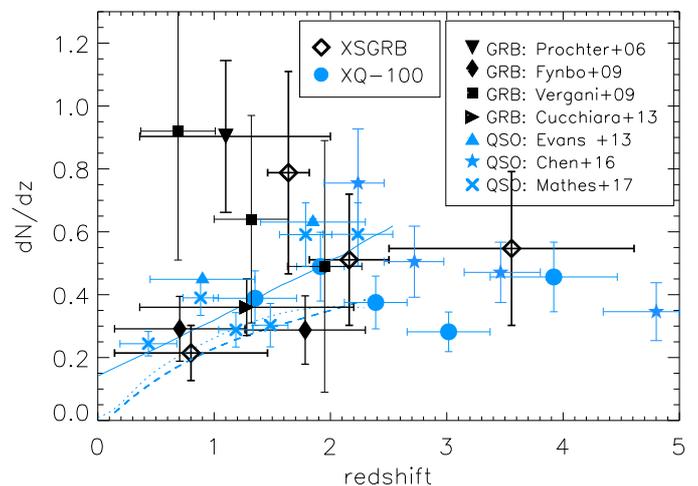}
  \caption{Incidences from the literature for both GRB and quasar
    absorbers. Symbols are colour coded with all quasar samples shown
    in blue and GRB samples in black. Incidences from the low spectral
    resolution GRB sample from \citet{fynbo09} were re-analysed in
    this work. This low resolution data-set is also part of the
    sample compiled by \citet{cucchiara13}.  The dashed and dotted
    curves show expectations from SDSS quasar data from
    \citet{prochter06q} and \citet{zhu13}, respectively, and the solid
    line the incidence towards quasars measured with high spectral
    resolution data \citep{mathes17}.}
\label{fig:line_dens2}
\end{figure}

Whereas the incidences from the two legacy samples are consistent with
each other, we proceed to compare incidences to other samples of
intervening strong \mgii absorbers towards GRB reported in the
literature and to SDSS quasars. In Fig.~\ref{fig:line_dens2} all GRB
samples have black symbols and quasar samples are shown in blue. The
large black diamonds that represent the XSGRB sample show that the
incidence rate from GRBs at $0.144<z<1.449$ agrees with incidence rate
from SDSS quasars illustrated by the dashed and dotted curves
\citep{prochter06q,zhu13}, while the bin for GRBs at $1.459<z<1.818$
suggests a larger incidence of $dN/dz=0.788\pm0.322$
(Table~\ref{tab:dndz}) which is an excess factor of 2.3$\pm$0.9
compared to $dN/dz=0.354\pm0.005$ expected from the SDSS
\citep{zhu13}. To compute uncertainties for the SDSS incidence we fit
$dN/dz$ \citep[Fig. 13 in][]{zhu13} with a 3.rd order polynomial
function following the results by \citet{prochter06q} and calculated
the covariance matrix from which we derived 68\% confidence intervals.

In Fig.~\ref{fig:line_dens2} the points at $z\sim1$ from
\citet{prochter06g} and at $z\sim0.7$ from
\citet{vergani09}\footnote{\citet{bergeron11} present a correction to
  the numbers given by \citet{vergani09} and we use the corrected
  incidences here.}, suggest incidences that have an excess of a
factor of 3.8 and 2, although consistent to within 1.5--2$\sigma$
uncertainties from the SDSS.  \citet{cucchiara13} determined an
average incidence $dN/dz = 0.18 \pm 0.06$ in a broad redshift range
$0.36<z<2.2$ based on their own independently collected sample of 83
GRBs. Their full sample consisted of 95 high- and low-resolution
afterglow spectra and included previously published data from the
literature which gave a higher incidence of $0.36\pm0.09$. Compared to
$dN/dz=0.24$ expected from SDSS quasars they derived an excess factor
of $1.5\pm0.4$. In the same redshift interval for GRB absorbers we
derived $dN/dz(z=0.36-2.2) = 0.42\pm0.10$, i.e. consistent within
0.5$\sigma$ uncertainties incidences from the total sample in
\citet{cucchiara13}.

We examined if any \mgii absorbers were potentially missed in the SDSS
samples by cross matching SDSS \mgii absorbers in the study by
\citet{raghunathan16} with the XQ-100 sample. In total 17 quasars are
in common, and for these sight lines the detected strong absorbers
mostly agree. One of the SDSS spectra suggests $W_r$ below the 1~{\AA}
criterion, whereas in the XQ-100 data the criterion is fulfilled, and
reversely for one other case. Only a single strong \mgii system at
$z\sim2$ was missed in the SDSS data (SDSS J105705+191041). The number
of objects in common in XQ-100 and in \citet{raghunathan16} is
insufficient to demonstrate if the SDSS is missing any absorbers since
XQ-100 does not probe absorbers below $z=1$. More detailed
investigations including lower redshift quasars are needed to
determine if other absorbers could be missed in the SDSS.

\subsection{Comparison with high spectral resolution quasar data}
While the SDSS database of quasar spectra provides the largest sample
for comparison in terms of number statistics, other large samples of
quasar spectra have been compiled. Using archival high spectral
resolution data from Keck/HIRES and VLT/UVES, \citet{evans13}
investigated \mgii systems in 252 quasar spectra, and determined an
incidence of strong absorbers $dN/dz=0.449\pm0.003$ at $z<1.4$,
consistent with our result within 1$\sigma$ errors, but above the
average SDSS incidence ($dN/dz(z=0.43-1.4)=0.256\pm0.003$) by a factor
of $1.75\pm0.02$. They also found that more luminous quasars at
$0.4<z<2.3$ have statistically significant fewer absorbers than
fainter quasars, but the difference in incidence rates is only a 20\%
effect. In a compiled sample of 602 high-resolution quasar spectra,
again obtained from archival UVES and HIRES data, \citet{mathes17}
analysed \mgii absorbers at $0.14<z<2.64$ (blue crosses in
Fig.~\ref{fig:line_dens2}) and found increasing incidences at $z\sim2$
compared to lower redshifts, generally consistent with the redshift
evolution of absorbers seen in SDSS spectra, although the incidences
found by \citet{mathes17} are 2$\sigma$ higher at redshift bins
$z<1.1$ and $z>1.5$.

While we computed an excess of strong \mgii absorbers relative to the
SDSS quasars, the incidences of strong absorbers from our two
X-shooter samples are consistent with each other to within 0.3$\sigma$
uncertainties.  Using instead $\frac{\partial N}{\partial z}$ from the
higher spectral resolution quasar data parametrized by
\citet{mathes17} to compute the expected numbers of absorbers from
equation~\ref{eq:nabs}, we would expect 16.2 and 50.9 strong
absorbers, resulting in no evidence for any excess with factors of
$1.11\pm0.29$ (GRBs) and $1.02\pm0.18$ (quasars), respectively.

There are several possible causes for a discrepancy between the \mgii
absorber incidences in high-resolution data and the SDSS.  Choosing to
use any available high-resolution quasar spectra in the archives may
cause a bias.  Particularly, any observing programmes aimed at
follow-up DLA studies will force $dN/dz$ to become artificially high
for strong \mgii absorbers, while no bias is expected for weak
absorbers as weak systems are rarely targeted on purpose.  Other
observing programmes may have targeted quasars specifically selected
to not have strong intervening absorbers. To what extent these
selections affect the reported incidences in the literature is
unclear.  \citet{mathes17} noted a slight excess of strong \mgii
absorbers compared to SDSS, specifically for the strongest ones with
$W_r>3$~{\AA}.  They performed K-S tests on similarly sized samples of
weak plus strong absorbers ($W_r>0.3$ {\AA}) drawn from SDSS quasars
and by comparing the equivalent width distribution functions $f(W_r,
W^*)= (N^*/W^*)exp(-W_0/W^*)$ they found similar characteristic
equivalent widths, $W^*$ and concluded that their high-resolution
samples were not biased. They did not compare the strong absorbers
alone, nor did they investigate discrepancies in incidences. In a
similar manner, \citet{evans13} compared the equivalent distribution
function of strong \mgii absorbers with that of SDSS absorbers studied
by \citet{nestor05}, and by performing K-S tests they concluded that
their samples were not inconsistent within a 2$\sigma$ level to the
SDSS.  This is surprising as the incidence data points in
Fig.~\ref{fig:line_dens2} are inconsistent with the SDSS.

We speculate that the automated searches for \mgii absorbers in SDSS
may miss some of the absorbers even though the search algorithms take
into account S/N ratios and detection limits of the absorbers as well
as correcting for incompleteness. However, because different
publications of incidences based on SDSS quasars largely agree, we do
not consider this a valid explanation.  In this work, we compared the
incidence to SDSS incidences by \citet{zhu13}, but note that other
authors have measured slightly different incidences from various SDSS
data releases. For example, \citet{seyffert13} reported
$dN/dz(z=0.36-2.28)= 0.293\pm0.002$ while the incidence in this
redshift interval in the study by \citet{zhu13} was $dN/dz=
0.306^{+0.001}_{-0.005}$ and \citet{nestor05} reported $dN/dz=
0.278\pm0.010$.  If we chose other references for SDSS incidences, the
excess in high-resolution data would have been slightly
higher. Nevertheless, the variations between incidences derived from
various analysis methods and SDSS data releases are small compared to
the excess seen for the high-resolution data.

A third plausible cause for discrepancies is the quasar optical color
selection criteria adopted by SDSS, which is biased against red, dusty
quasars \citep{krogager16,fynbo17}. Some of these quasars are reddened
by dust in intervening DLAs or by dust in additional strong
intervening \mgii absorbers (Heintz et al. 2017 in prep.).  Heintz et
al. estimate that about 10\% of quasars are missed by SDSS or the
Baryonic Oscillations Spectroscopic Survey (BOSS), whereas
\citet{menard08} estimate that only 1\% are missed by dust in
foreground \mgii absorbers with $W_r=1$~{\AA} but with a percentage
that increases to 50\% for the strongest absorbers ($W_r=6$~{\AA}).
\citet{raghunathan16} find that about 10\% of the 266,433 quasars in
SDSS DR12 contain strong \mgii absorbers. If a very large fraction of
the missed quasars have strong intervening absorbers, the true \mgii
absorber incidence towards quasars could be a factor of 2 higher than
seen in SDSS, thereby explaining the excess we see in high-resolution
data.  The factor of 2 higher incidence is an upper limit because
quasars may be reddened by their hosts rather than by intervening
absorption systems \citep{krogager15,krogager16}.  To determine if
there is a significant excess of strong \mgii absorbers towards these
red quasars and measure the effect of the incidences on SDSS quasar
selection biases, requires that the incidence and equivalent
distribution function of intervening absorbers towards the reddened
quasars are analysed. This will be the focus of future work.

Some quasars in the XQ-100 survey are selected from SDSS, so any
potential selection biases in SDSS would propagate into this survey,
as well as other high-spectral resolution quasar studies.
Radio-selected quasar samples would not propagate this dust bias, but
a comparison of \mgii absorber incidences towards radio- and optically
selected quasars did not show any difference \citep{ellison04}. The
search for strong \mgii absorbers towards blazars \citep{bergeron11}
included both radio- and optically (SDSS) selected targets, but with 9
optically selected and very bright targets from a sample of 45
objects, any bias carried by SDSS would be small, and could also cause
an excess of \mgii towards blazars like those seen in high-resolution
quasars studies.

\subsection{Comparison at higher redshifts, $z\gtrsim2.3$}
Because the SDSS spectra only cover visible wavelengths, the \mgii
samples extend to only $z\approx2.3$. Expanding the redshift range by
observing higher redshift quasars with near-IR spectra at an
intermediate resolution of $R\sim6000$ with Magellan/FIRE,
\citet{matejek12} showed that beyond this redshift, the incidence rate
is constant within the uncertainties ($dN/dz\sim 0.575-0.325$) out to
$z=5.35$. These values are consistent with our computed incidence
rates beyond $z\gtrsim2$. With an expanded sample of 100 quasars at
$3.55<z<7.08$ observed with Magellan/FIRE, \citet{chen16} found
evidence for a redshift evolution and a declining incidence from $z=2$
to $z=6$, as illustrated by the star symbols in
Figures~\ref{fig:line_dens} and ~\ref{fig:line_dens2}.

We computed the expected number of strong systems from the
parametrization of \citet{chen16} using Equation~\ref{eq:nabs} to be
7$\pm$1 and 81$\pm$74, whereas we found 4 and 45 towards GRBs and
quasars, respectively.  This gives excess factors of $0.71\pm0.34$ and
$0.56\pm0.93$, as reported in Table~\ref{tab:literature}.  Even though
our sample of GRBs is larger than in previous studies, the redshift
path length ($\Delta z=13.14$ at $z>2.3$ in
Table~\ref{tab:literature}) is significantly smaller than path lengths
in other high-resolution quasar studies \citep{matejek12,chen16}, and
this causes a higher uncertainty of the measured incidences.

\subsection{Excess due to lensing?}
  \citet{vergani09} suggest lensing magnification of part of the UVES
  sample as a possible explanation of the excess, as high-resolution
  UVES spectra can be obtained only for bright afterglows. Such bright
  afterglows could have an increased probability of having strong
  \mgii intervening systems (causing the magnification) along lines of
  sight. Indeed, their sample includes one remarkable burst
  GRB~060418, which has three strong intervening \mgii systems, two of
  which lie at $z\sim0.6$ \citep{ellison06b}.  \citet{prochter06g}
  noted that excluding this remarkable GRB does not significantly
  change the statistics. An examination of the field with imaging and
  follow-up spectra of galaxies near to the GRB line of sight
  suggested some effects of gravitational magnification
  \citep{pollack09}.  However, adopting a different lensing
  configuration based on recent VLT/MUSE data of this field (PI:
  Schulze) with spectroscopic identifications of the host galaxies of
  the absorbers (Christensen \& Schulze in prep.) we find that any
  lensing magnification is very unlikely for this burst. To produce a
  significant gravitational magnification the nearby hosts would have
  to be $\sim$3 orders of magnitude more massive than those detected.

For another burst, GRB~030429, a similar conclusion was made that
strong gravitational lensing did not cause magnification, although an
intervening galaxy was found close to the GRB line of sight
\citep{jakobsson04}. On the case-by-case basis we still have no
evidence that GRBs are gravitationally magnified.

\subsection{The $W_r$ distribution}
\begin{figure}[t!]
\centering
\includegraphics[bb=77 366 533 693, clip, width=\linewidth]{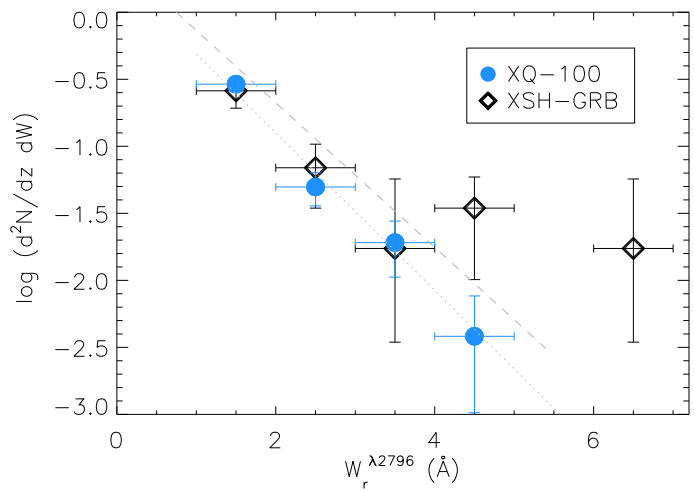}
 \caption{The equivalent width distribution of $W_r^{2796}$ of the GRB
   and quasar samples. Vertical error bars were computed using the
   $\sqrt{N_{\mathrm{abs}}}$ approximation. The dotted and dashed lines
   represent best fits for the distribution of quasar absorbers by
   \citet{mathes17} and \citet{chen16}, respectively.}
\label{fig:ewdist}
\end{figure}

While we have found that the incidence rates agree between most
samples within the uncertainties, the analysis has not considered the
distribution of the absorber equivalent widths. We investigate the
rest-frame equivalent width distribution by computing the function:
\begin{equation}
  f(W) = \frac{d^2N}{dz dW}.
\end{equation}

Since our total \mgii sample sizes are not very large, we computed the
distribution within the full integrated redshift range. This choice
can be justified as \citet{chen16} found no significant redshift
evolution of the $W_r$ distribution for the strongest absorbers. The
results presented in Fig.~\ref{fig:ewdist} demonstrate a good overall
agreement between the GRB and quasar samples. Although very strong
absorbers with $W_r>4$ {\AA} appear to be more frequent towards GRBs
than quasars, the numbers agree within $\sim1\sigma$ uncertainty levels.
The only outlier is the highest $W_r$ point from the GRB absorbers,
which presents a single exceptionally strong \mgii absorber towards
GRB~120815A \citep{kruhler13}. However, the bin only contains a single
data point, and within the uncertainties the bin contains
$1^{+2.3}_{-0.8}$ absorbers according to Poisson statistics
\citep{gehrels86}, such that the frequency could be very low as well.

Compared to the $W_r$ distributions from quasar \mgii absorbers in the
literature \citep{chen16,mathes17}, although marginally consistent
within the errors, the absolute values of the GRB sample could suggest
a higher frequency of the strongest absorbers.
\citet{chen16} reported that the very strong absorbers in their quasar
sample were mainly found at $z>1.5$; a trend which we confirm in our
GRB and quasar data.

\section{Summary}
\label{sect:conc}
We combined two VLT/X-shooter legacy data sets consisting of quasar
and GRB afterglow spectra to explore the incidence rates of
intervening strong \mgii systems with $W_r^{\lambda2796}>1$ {\AA} out
to $z\sim5$. With the combination of uniformly selected data, we
have covered large redshift path lengths for both samples, although at
$z>2.3$, the quasar sample probes a ten times larger survey path than
the GRB sample. Based on number statistics of absorbers we find that:

\begin{itemize}
\item There is no discrepancy between the incidence of strong \mgii
  absorbers at $0.1<z<5$ between the two samples. Number statistics
  show agreements within 1$\sigma$ uncertainty levels at all redshift bins.
  
\item Our determined incidence of strong \mgii systems towards quasars
  agrees within 1$\sigma$ with other intermediate- to high spectral
  resolution ({\bf $R\gtrsim6000$}) quasar studies \citep{evans13,mathes17}
  at redshifts $0.14<z<2.6$.

\item The decline of the incidence of strong \mgii absorbers at higher
  redshifts ($2<z<6$) towards quasars \citep{chen16} is not seen in
  our sample. However, we can not exclude a decline beyond $z\sim2.5$
  due to large uncertainties of the incidences.  We do not detect any
  absorbers beyond $z\sim4.6$.
 
\item At lower redshifts ($0.14<z<2.3$), the incidence of strong \mgii
  absorbers in the X-shooter data suggests an overdensity of a factor
  of $1.5\pm0.4$ (GRBs) and $1.4\pm0.2$ (quasars) relative to the
  large statistical database of strong \mgii absorbers in SDSS quasars
  \citep{prochter06g,zhu13,raghunathan16}. The only remaining
  discrepancy in the incidence of strong \mgii absorbers is therefore
  between SDSS quasars and higher spectral resolution quasar
  studies. We suggest that this discrepancy can be explained by a
    selection bias in SDSS against quasars reddened by dust in
    intervening absorbers.

\item Incidences from the low-resolution ($R\lesssim 2000$) GRB
  afterglow spectra from FORS analysed in this work agree within their
  1$\sigma$ error bars with SDSS quasar data, but also with the
  X-shooter intermediate spectral resolution results due to the large
  error-bars.

\item The distribution of absorber equivalent widths with
  $1<W_r^{\lambda2796}<7$ {\AA} is consistent within 1$\sigma$
  uncertainty levels between the quasar and the GRB afterglow
  samples. The distributions are also consistent with measurements
  from high spectral resolution studies of quasars absorbers
  \citep{chen16,mathes17}.

\end{itemize}
  
The agreement between incidences towards GRB afterglows and quasars
demonstrates that both object types probe random lines of sight
through the universe. None of the objects are more likely than the
other to encounter an intervening strong \mgii absorber.  The measured
incidences and evolution with redshift of both types of line of sight
are therefore consistent with the notion that strong \mgii systems can
be used as un-biased tracers of the global star formation rate density
of the universe out to the highest redshifts
\citep{menard11}. Detecting very high redshift GRBs with intervening
very high redshift \mgii absorbers will therefore provide a useful
tracer of the star formation density at the earliest epochs in the
universe.

\begin{acknowledgements}
We acknowledge valuable comments from an anonymous referee.

LC and RC are supported by YDUN grant DFF  4090-00079.

SDV is supported by the French National Research Agency (ANR) under
contract ANR-16-CE31-0003 BEaPro.

SL has been supported by FONDECYT grant number 1140838 and partially
by PFB-06 CATA.

AdUP and CT acknowledge support from Ram\'on y Cajal fellowships, and
with RSR a BBVA Foundation Grant for Researchers and Cultural
Creators, and the Spanish Ministry of Economy and Competitiveness
through project AYA2014-58381-P.

GB acknowledges support from the National Science Foundation through
grant AST-1615814.

ZC acknowledges support from the Juan de la Cierva Incorporaci\'on
fellowship IJCI-2014-21669 and from the Spanish research project AYA
2014-58381-P.

PJ and KEH acknowledge support by a Project Grant (162948–051) from
The Icelandic Research Fund.

JJ acknowledges support from NOVA and NWO-FAPESP grant for
    advanced instrumentation in astronomy.

RSR acknowledges support from ASI (Italian Space Agency) through the
Contract n. 2015-046-R.0 and from European Union Horizon 2020
Programme under the AHEAD project (grant agreement n. 654215).

MV gratefully acknowledges financial support from the Danish Council
for Independent Research via grant no. DFF 4002-00275.

\end{acknowledgements}


\bibliography{paper}
\clearpage
\input{qso_abslist2}

\end{document}

%% file: qso_abslist2.tex
\begin{table*}
\caption{Intervening absorbers in the QSO sample. Quasar and strong
  \mgii system redshifts are listed as $z_{\mathrm{QSO}}$ and
  $z_{\mathrm{abs}}$. }
\centering
  \begin{multicols}{2}

  \begin{supertabular}{llll}
    \hline \hline
    QSO name & $z_{\mathrm{QSO}}$ & $z_{\mathrm{abs}}$ & $W_r^{\lambda2796}$ [{\AA}] \\
    \hline
J0003--2603 & 4.12 &  3.3899  &  1.318$\pm$0.012 \\
J0006--6208 & 4.44 &  1.9580  &  1.845$\pm$0.021 \\
 	   &  	  &  3.7762  &  1.100$\pm$0.029 \\
J0034+1639 & 4.29 &  1.7997  &  1.556$\pm$0.016 \\
 	   &  	  &  2.8524  &  1.270$\pm$0.064 \\
J0042--1020 & 3.86 &  2.7545  &  1.929$\pm$0.018 \\
 	   &  	  &  3.6286  &  1.158$\pm$0.027 \\
J0056--2808 & 3.63 &  1.3411  &  1.221$\pm$0.020 \\
J0100--2708 & 3.55 &  2.1484  &  1.366$\pm$0.019 \\
J0113--2803 & 4.31 &  1.3898  &  1.506$\pm$0.017 \\
           &      &  3.0160  &  4.000$\pm$0.066 \\
J0117+1552 & 4.24 &  2.5227  &  3.041$\pm$0.029 \\
J0124+0044 & 3.84 &  2.2609  &  1.603$\pm$0.015 \\
J0134+0400 & 4.18 &  1.6656  &  1.637$\pm$0.010 \\
 	   & 	  &  3.7725  &  2.287$\pm$0.014 \\
J0137--4224 & 3.97 &  3.1010  &  1.013$\pm$0.096 \\
J0153--0011 & 4.19 &  1.9080  &  1.126$\pm$0.040 \\
J0211+1107 & 3.97 &  1.4689  &  1.780$\pm$0.022 \\
           &      &  2.5250  &  1.318$\pm$0.036 \\
           & 	  &  3.1415  &  1.533$\pm$0.031 \\
 	   & 	  &  3.5033  &  1.685$\pm$0.044 \\
J0214--0518 & 3.98 &  1.3071  &  1.075$\pm$0.013 \\
 	   & 	  &  1.5279  &  2.901$\pm$0.018 \\
J0234--1806 & 4.30 &  3.6934  &  2.182$\pm$0.053 \\
           &      &  4.2280  &  1.877$\pm$0.047 \\
J0244--0134 & 4.05 &  2.1011  &  1.688$\pm$0.016 \\
J0247--0555 & 4.23 &  1.7099  &  1.176$\pm$0.019 \\
J0255+0048 & 4.00 &  3.2551  &  2.576$\pm$0.032 \\
	   &  	  &  3.4502  &  0.972$\pm$0.027 \\
J0307--4945 & 4.72 &  2.6288  &  1.047$\pm$0.036 \\
 	   &  	  &  3.5909  &  1.379$\pm$0.015 \\
           &      &  4.2110  &  2.038$\pm$0.020 \\
           &  	  &  4.4658  &  1.665$\pm$0.019 \\
J0311--1722 & 4.03 &  1.9408  &  1.321$\pm$0.014 \\
J0415--4357 & 4.07 &  3.8076  &  2.102$\pm$0.024 \\
J0424--2209 & 4.33 &  4.1418  &  0.983$\pm$0.023 \\
J0523--3345 & 4.39 &  1.5705  &  1.056$\pm$0.010 \\
J0529--3526 & 4.42 &  2.1907  &  1.588$\pm$0.041 \\
J0747+2739 & 4.13 &  2.6100  &  2.435$\pm$0.184 \\
           &      &  3.4221  &  1.511$\pm$0.025 \\
J0818+0958 & 3.66 &  2.8270  &  1.124$\pm$0.033 \\
	   &  	  &  3.3061  &  1.532$\pm$0.028 \\
J0833+0959 & 3.72 &  1.2911  &  2.224$\pm$0.014 \\
J0835+0650 & 4.01 &  1.5096  &  1.727$\pm$0.018 \\
           & 	  &  3.1894  &  1.181$\pm$0.036 \\
J0839+0318 & 4.23 &  2.1340  &  1.375$\pm$0.065 \\
J0920+0725 & 3.65 &  1.5790  &  1.206$\pm$0.010 \\
    	   &      &  2.1437  &  1.266$\pm$0.012 \\
    	   &      &  2.2368  &  1.331$\pm$0.012 \\
J0935+0022 & 3.75 &  1.2827  &  2.787$\pm$0.027 \\
J0937+0828 & 3.70 &  2.1440  &  2.260$\pm$0.021 \\
J0955--0130 & 4.42 &  2.6235  &  1.859$\pm$0.123 \\
J0959+1312 & 4.09 &  1.8727  &  2.850$\pm$0.012 \\
J1020+0922 & 3.64 &  1.8055  &  1.110$\pm$0.023 \\
J1032+0927 & 3.98 &  2.2617  &  0.997$\pm$0.017 \\
J1034+1102 & 4.27 &  1.6004  &  1.992$\pm$0.012 \\
    	   &      &  2.1167  &  2.377$\pm$0.013 \\
J1042+1957 & 3.63 &  2.1424  &  1.052$\pm$0.018 \\
J1054+0215 & 3.97 &  1.4944  &  1.616$\pm$0.025 \\
J1057+1910 & 4.13 &  1.9853  &  1.285$\pm$0.044 \\
    	   &      &  3.3735  &  2.361$\pm$0.095 \\
J1058+1245 & 4.34 &  3.4305  &  1.501$\pm$0.038 \\
   	   &      &  2.1097  &  3.079$\pm$0.019 \\
       	   &      &  2.1826  &  1.569$\pm$0.025 \\
J1103+1004 & 3.61 &  1.7543  &  1.148$\pm$0.015 \\
J1108+1209 & 3.68 &  1.8693  &  2.376$\pm$0.019 \\
   	   &      &  3.5450  &  1.761$\pm$0.025 \\
J1110+0244 & 4.14 &  2.1199  &  3.021$\pm$0.015 \\
J1111--0804 & 3.92 &  1.9769  &  1.798$\pm$0.015 \\
J1201+1206 & 3.52 &  1.9973  &  1.541$\pm$0.007 \\
J1202--0054 & 3.59 &  2.7696  &  0.986$\pm$0.036 \\
J1249--0159 & 3.63 &  3.1020  &  1.786$\pm$0.022 \\
J1304+0239 & 3.65 &  1.8075  &  1.330$\pm$0.013 \\
   	   &      &  3.2108  &  4.699$\pm$0.023 \\
J1312+0841 & 3.73 &  1.9183  &  1.155$\pm$0.013 \\
    	   &      &  2.6594  &  1.138$\pm$0.054 \\
J1320--0523 & 3.72 &  1.4038  &  1.969$\pm$0.013 \\
   	   &      &  1.5277  &  1.488$\pm$0.012 \\
J1323+1405 & 4.05 &  1.4974  &  1.048$\pm$0.020 \\
   	   &      &  1.9824  &  1.634$\pm$0.041 \\
J1330--2522 & 3.95 &  3.0810  &  1.426$\pm$0.019 \\
J1331+1015 & 3.85 &  1.5793  &  1.122$\pm$0.016 \\
J1352+1303 & 3.71 &  3.0075  &  2.106$\pm$0.024 \\
J1416+1811 & 3.59 &  2.1150  &  1.145$\pm$0.018 \\
           &      &  2.2277  &  2.449$\pm$0.019 \\
J1421+0643 & 3.69 &  1.4580  &  1.539$\pm$0.014 \\
           &      &  2.6158  &  1.178$\pm$0.073 \\
J1442+0920 & 3.53 &  1.1233  &  1.048$\pm$0.013 \\
J1524+2123 & 3.60 &  2.5449  &  0.997$\pm$0.133 \\
J1542+0955 & 3.99 &  3.2818  &  1.091$\pm$0.056 \\
J1552+1005 & 3.72 &  3.6665  &  1.772$\pm$0.037 \\
J1621--0042 & 3.71 &  1.1341  &  3.167$\pm$0.016 \\
           &      &  3.1050  &  1.504$\pm$0.020 \\
J1633+1411 & 4.36 &  2.2345  &  1.007$\pm$0.015 \\
J1723+2243 & 4.53 &  3.6958  &  3.762$\pm$0.018 \\
J2215--1611 & 3.99 &  2.3910  &  1.284$\pm$0.018 \\
J2216--6714 & 4.48 &  2.0615  &  1.455$\pm$0.011 \\
J2239--0522 & 4.55 &  3.0310  &  1.537$\pm$0.020 \\
J2251--1227 & 4.16 &  2.8826  &  1.796$\pm$0.039 \\
J2344+0342 & 4.25 &  2.5541  &  1.156$\pm$0.040 \\
    	   &      &  3.2201  &  1.458$\pm$0.018 \\
J2349--3712 & 4.22 &  2.8305  &  1.477$\pm$0.064 \\
\hline
  \end{supertabular}
  \end{multicols}
\label{tab:list_qso}

\end{table*}